\documentclass[twocolumn,superscriptaddress,amsmath,amssymb,prl,longbibliography]{revtex4-1}

\usepackage{amssymb,amsmath}
\DeclareMathAlphabet\mathbfcal{OMS}{cmsy}{b}{n}
\usepackage{cmap}
\usepackage{graphicx} 
\usepackage{bm}
\usepackage{color}
\usepackage{blindtext}
\usepackage{hyperref}
\usepackage{listings}
\usepackage{booktabs}
\usepackage{multirow}
\usepackage{amsmath}
\usepackage{mathrsfs}
\usepackage{braket}
\usepackage{mathtools}

\usepackage{dsfont}
\usepackage{tabularx}

\newcommand{\blue}[1]{{\color{black}{#1}}}

\begin{document}

\title{Water-Wave Vortices and Skyrmions}

\author{Daria A. Smirnova}
\affiliation{Theoretical Quantum Physics Laboratory, Cluster for Pioneering Research, RIKEN, Wako-shi, Saitama 351-0198, Japan}
\affiliation{
Research School of Physics, Australian National University, Canberra, ACT 2601, Australia}

\author{Franco Nori}
\affiliation{Theoretical Quantum Physics Laboratory, Cluster for Pioneering Research, RIKEN, Wako-shi, Saitama 351-0198, Japan}
\affiliation{Center for Quantum Computing (RQC), RIKEN, Wako-shi, Saitama 351-0198, Japan}
\affiliation{Physics Department, University of Michigan, Ann Arbor, MI 48109-1040, USA}

\author{Konstantin Y. Bliokh}
\affiliation{Theoretical Quantum Physics Laboratory, Cluster for Pioneering Research, RIKEN, Wako-shi, Saitama 351-0198, Japan}
\affiliation{Centre of Excellence ENSEMBLE3 Sp. z o.o., 01-919 Warsaw, Poland}
\affiliation{Donostia International Physics Center (DIPC), Donostia-San Sebasti\'{a}n 20018, Spain}

\begin{abstract}
Topological wave structures -- phase vortices, skyrmions, merons, etc. -- are attracting enormous attention in a variety of quantum and classical wave fields. Surprisingly, these structures have never been properly explored in the most obvious example of classical waves: water-surface (\blue{gravity-capillary}) waves. Here we fill this gap and describe: (i) water-wave vortices of different orders carrying quantized angular momentum with orbital and spin contributions, (ii) skyrmion lattices formed by the instantaneous displacements of the water-surface particles in wave interference, (iii) meron (half-skyrmion) lattices formed by the spin density vectors, as well as (iv) spatiotemporal water-wave vortices and skyrmions. We show that all these topological entities can be readily generated in linear water-wave interference experiments. Our findings can find applications in microfluidics and show that water waves can be employed as an attainable    playground for emulating universal topological wave phenomena.
\end{abstract}

\maketitle

{\it Introduction.---}Wave vortices are universal physical entities with nontrivial topological and dynamical properties: quantized phase increments around point phase singularities and quantum-like angular momentum (AM). Examples of wave vortices are known since the 19th century, these has been observed and explored in tidal \cite{Nye1988}, quantum-fluid \cite{Yarmchuk1979,Fetter2009}, optical  \cite{Berry1981,Allen_book,Soskin2001}, sound \cite{Hefner1999,Volke2008,Guo2022}, elastic \cite{Chaplain2022_II}, surface-plasmon \cite{Gorodetski2008,Prinz2023}, exciton-polariton \cite{Dall2014}, quantum electron \cite{Bliokh2017}, neutron \cite{Clark2015}, and atom \cite{Luski2021} waves.

Strikingly, wave vortices have not been properly studied in the most obvious example of classical waves: water-surface (\blue{gravity-capillary}) waves. Only a recent series of experiments \cite{Filatov2016,Filatov2016_II,Francois2017,Bliokh2022} described the generation of a square lattice of alternating vortices in the interference of orthogonal standing water waves. 

However, the theoretical description of these experiments lacks the identification with {\it wave vortices}, very different from the usual hydrodynamical vortices. It was indicated that the hydrodynamical vorticity appears due to nonlinearity \cite{Filatov2016,Filatov2016_II}, and that these vortices are closely related to the Stokes drift and AM \cite{Francois2017,Bliokh2022}, but no quantized topological and dynamical properties have been indicated. Furthermore, only the simplest first-order vortices were produced (cf., e.g., quantum-electron vortices of higher orders $\sim 10^2 \!-\!10^3$ \cite{McMorran2011,Tavabi2022}).

In this work, we describe water-wave vortices (WWVs) in \blue{gravity-capillary} waves. We reveal their topological properties and show that circularly-symmetric vortices are eigenmodes of the {\it total AM} operator, including the spin and orbital parts. In the linear approximation, WWVs have {\it zero vorticity}. Nonetheless, the quadratic {\it Stokes drift} produces slow orbital motion of water particles and nonzero nonlinear vorticity. Importantly, water particles experience two kinds of circular motions with different spatial and temporal scales: (i) local linear-amplitude-scale circular motion with the wave frequency in the linear regime and (ii) slow wavelength-scale circular motion due to the nonlinear Stokes drift. These two motions are responsible for the spin and orbital contributions to the quantized total AM. 

Moreover, water waves have inherent {\it vector} properties: the local Eulerian displacement of water-surface particles is a counterpart of the 3D polarization in optical or acoustic wavefields \cite{Sugic2020,Bliokh2021POF}. Therefore, following great recent progress in the generation of topological vector entities -- {\it skyrmions} \cite{Nagaosa2013} -- in classical electromagnetic \cite{Tsesses2018,Du2019,Dai2020,Gao2020,Shen2021,Lei2021,Deng2022}, sound \cite{Ge2021,Muelas2022}, and elastic \cite{Cao2023} waves, here we describe {\it water-wave skyrmions}. We show that the interference of three plane water waves can generate a hexagonal lattice of: (i) WWVs; (ii) skyrmions of the instantaneous water-particle displacements and (iii) {\it merons} (half-skyrmions) of the local spin density. This field configuration is just one step from the recent experiments \cite{Filatov2016,Filatov2016_II,Francois2017,Bliokh2022}, and is quite feasible for the experimental implementation.

Finally, following enormous current interest in {\it spacetime} structured waves \cite{Yessenov2022,Shen2023}, in particular {\it spatiotemporal vortices} \cite{Sukhorukov2005,Bliokh2012,Hancock2019,Chong2020,Bliokh2021}, we show that detuning the frequency of one of the interfering waves, one can readily produce moving lattices of spatiotemporal WWVs and {\it spatiotemporal skyrmions}.   

Thus, we reveal new structures with remarkable topological and dynamical properties in linear water waves. We argue that water waves offer a perfect classical platform for emulating universal quantum and topological wave phenomena, which can also find useful applications in microfluidics \cite{Ding2013,Wu2019}. 

\begin{figure*}[t]
\centering
\includegraphics[width=0.9\linewidth]{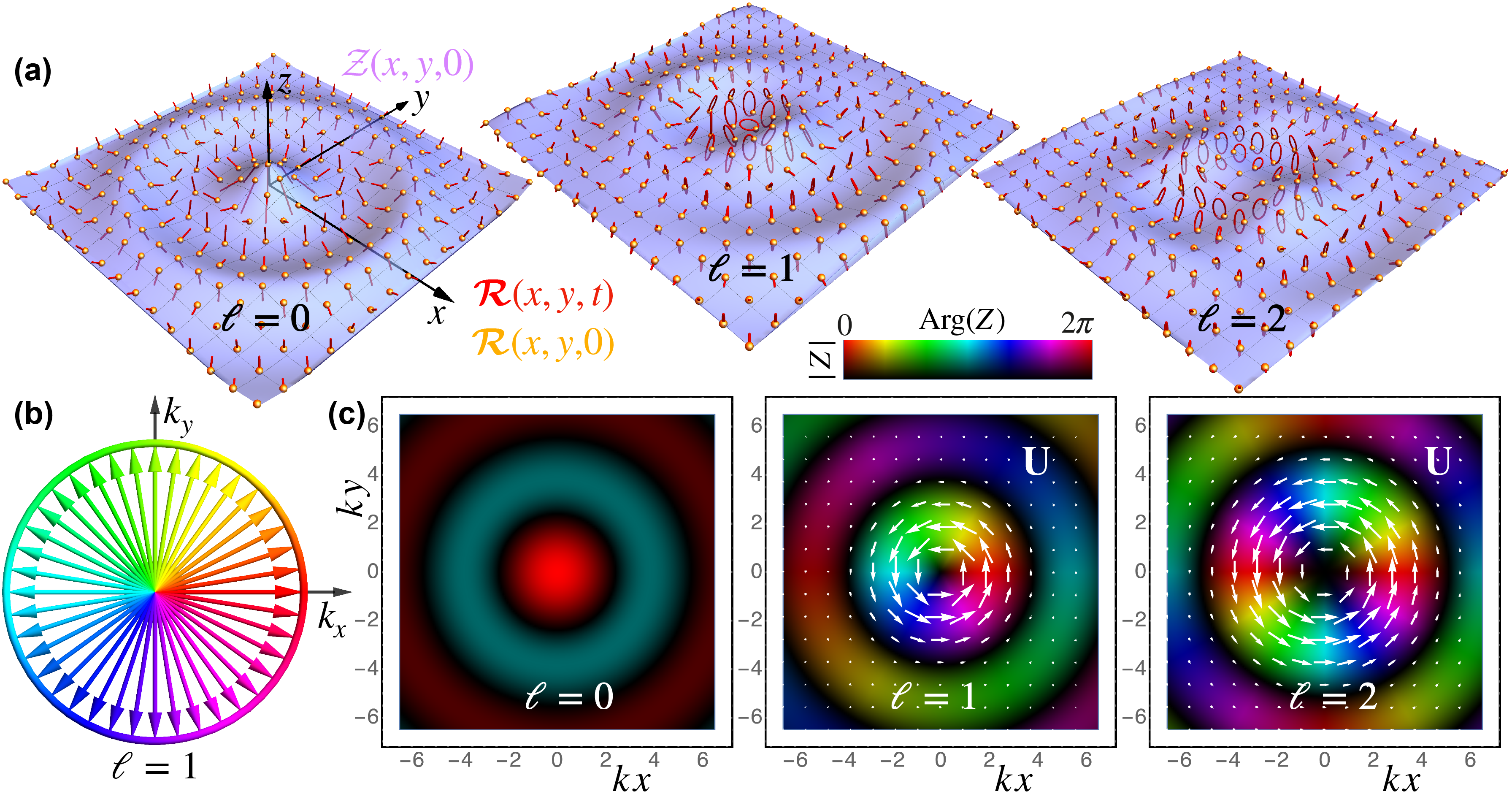}
\caption{(a) Instantaneous water surfaces ${\mathcal Z}(x,y,0)$ and Eulerian water-surface particle trajectories $\mathbfcal{R}(x,y,t)$ for circular WWVs with different topological charges $\ell$, Eqs.~(\ref{Z_Bessel}) and (\ref{a_Bessel}). The spin density ${\bf S}$ is directed normally to the elliptical particle trajectories and quantifies the AM of this elliptical motion. (b) The plane-wave spectrum of a circular WWV with color-coded phases for $\ell =1$. 
(c) The complex vertical-displacement field $Z(x,y)$ for WWVs from panel (a), with the phases and amplitudes coded by the colors and brightness, respectively. The white arrows indicate the second-order Stokes drift ${\bf U}$, Eq.~(\ref{Stokes}), characterizing the wave momentum density.}
\label{Fig_1}
\end{figure*}

{\it Water-wave vortices.---}We first consider monochromatic \blue{gravity-capillary} waves on a deep-water surface. The 3D Eulerian displacement of the water particles \blue{from} the $z=0$ surface is $\mathbfcal{R}({\bf r}_2,t) = {\rm Re}[{\bf R}({\bf r}_2)e^{-i\omega t}] = (\mathcal{X},\mathcal{Y},\mathcal{Z})$, where ${\bf R} = (X,Y,Z)$ is the complex displacement wavefield, ${\bf r}_2 = (x,y)$ and $\omega$ is the frequency. Separating the vertical and in-plane components of 3-vectors as ${\bf a} \equiv (a_x,a_y,a_z) = ({\bf a}_2, a_z)$, the wave equations of motion can be written as \cite{Bliokh2022,SM}
\begin{equation}
\label{EOM}
\blue{\omega^2 {\bf R}_2 \!= \!\left( g -\frac{\alpha}{\rho}\Delta_2 \right)\! {\bm \nabla}_2 Z,~
\omega^2 Z \!= \!-\! \left( g -\frac{\alpha}{\rho}\Delta_2 \right)\! {\bm \nabla}_2 \cdot {\bf R}_2.}
 \end{equation}
Here $g$ is the gravitational acceleration, \blue{$\alpha$ is the surface-tension coefficient, $\rho$ is the water density, $\Delta_2 = {\bm \nabla}_2 \cdot {\bm \nabla}_2$}, and Eqs.~(\ref{EOM}) \blue{with the plane-wave ansatz ${\bm \nabla}_2 \to i{\bf k}$ (${\bf k}$ is the wave vector)} yield the dispersion relation \blue{$\omega^2 = gk + (\alpha/\rho)k^3$}.

The vortex solutions of Eqs.~(\ref{EOM}) are obtained as a superposition of plane waves with wavevectors uniformly distributed along the $k_x^2 +k_y^2 = k^2$ circle with the azimuthal phase increment $2\pi \ell$, $\ell \in {\mathbb Z}$, Fig.~1(b). Constructing the complex vertical displacement in this way, we obtain:
\begin{equation}
\label{Z_Bessel}
Z = \frac{A}{2\pi} \int_0^{2\pi} e^{i {\bf k} \cdot {\bf r}_2 + i\ell \phi} d\phi =
A J_\ell (kr) e^{i\ell \varphi}\,.
 \end{equation}
Here $A$ is \blue{a constant} wave amplitude, $J_\ell$ is the Bessel function of the first kind, $\phi$ is the azimuthal angle in the $(k_x,k_y)$ plane, whereas $(r,\varphi)$ are the polar coordinates in the $(x,y)$-plane.

Equation (\ref{Z_Bessel}) describes 2D scalar cylindrical Bessel waves, Fig.~\ref{Fig_1}(c). However, water waves have a vectorial nature, and the other two components of the wavefield can be found from the first Eq.~(\ref{EOM}). It is convenient to write these in the basis of `circular polarizations' \cite{Picardi2018,Bliokh2022PRL}:
\begin{equation}
\label{a_Bessel}
R^\pm \equiv \frac{X\mp i Y}{\sqrt{2}} 
= \pm \frac{A}{\sqrt{2}} J_{\ell \mp 1}(kr)\, e^{i(\ell \mp 1)\varphi}.
 \end{equation}

In this basis, the $z$-component of the spin-1 operator, universal for classical vector waves, reads $\hat{S}_z = {\rm diag}(1,-1,0)$, while the  $z$-component of the orbital AM (OAM) operator is $\hat{L}_z = -i\partial_\varphi$ \cite{Allen_book}. Introducing the `wavefunction' $| \psi \rangle = (R^+, R^-, Z)$, one can see that the WWVs (\ref{Z_Bessel}) and (\ref{a_Bessel}), are {\it not} the OAM eigenmodes, but these are eigenomodes of \blue{the $z$-component of} the {\it total} AM with the quantized eigenvalue $\ell$:
\begin{equation}
\label{Jz}
\hat{J}_z | \psi \rangle = (\hat{L}_z + \hat{S}_z) | \psi \rangle = \ell\, | \psi \rangle\,.
\end{equation}
Such behavior (which can be interpreted as the inherent spin-orbit coupling) is a common feature of all cylindrical vector waves: optical \cite{Enk1994_II,Bliokh2010,Picardi2018}, quantum \cite{Bliokh2011PRL}, acoustic \cite{Bliokh2019_II}, and elastic \cite{Bliokh2022PRL}. 

Figure~\ref{Fig_1}(a) shows instantaneous water surfaces ${\mathcal Z}({\bf r}_2,0)$ and water-particle trajectories $\mathbfcal{R}({\bf r}_2,t)$ for WWVs with different $\ell$. The water-particle trajectories are 3D ellipses, entirely similar to the electric-field polarization in optical fields \cite{Bliokh2021POF}. The normal to the ellipse and its ellipticity determine the cycle-averaged AM \blue{of the particle}, i.e., {\it spin density} in water waves \cite{Jones1973,Bliokh2022}: ${\bf S} = (\rho \omega/ 2) {\rm Im}({\bf R}^*\! \times {\bf R})$. One can see that WWVs are characterized by inhomogeneous polarization textures. In the vortex center $r=0$, the polarization is purely vertical, $| \psi \rangle \propto (0,0,1)$, for $\ell =0$; it is purely circular, $| \psi \rangle \propto (1,0,0)$ and $| \psi \rangle \propto (0,1,0)$, for $\ell =\pm 1$; and the vector wavefield vanishes, $| \psi \rangle \propto (0,0,0)$, for $|\ell |>1$ \blue{(the vanishing of all vector wavefield components requires a higher-order degeneracy \cite{Picardi2018,Bliokh2022PRL,Bliokh2010,Bliokh2011PRL,Bliokh2019_II,Vernon2023})}. 

Importantly, WWVs are {\it not} the usual hydrodynamical vortices, which are formed by steady water motion with a nonzero circulation of the velocity $\mathbfcal{V} = \partial_t \mathbfcal{R}$ and vorticity ${\bm \nabla} \times \mathbfcal{V} \neq {\bf 0}$ \cite{LLfluid}. In contrast, linear monochromatic \blue{gravity-capillary} waves have zero vorticity: ${\bm \nabla} \times {\bf V} = {\bf 0}$, where ${\bf V} = -i\omega {\bf R}$ is the complex velocity field. This follows from Eqs.~(\ref{EOM}) and the incompressibility equation ${\bm \nabla} \cdot {\bf V} =0$. Wave vortices are {\it topological} entities with {\it quantized phase singularities} in the center. The `topological charge' can be defined in two equivalent ways \cite{BerryDennis2001,BAD2019}:   
\begin{equation}
\label{ell}
\frac{1}{2\pi}\!\oint {\bm \nabla}_2{\rm Arg}(Z) \cdot d{\bf r}_2 = \frac{1}{4\pi}\! \oint {\bm \nabla}_2{\rm Arg}({\bf R}\cdot {\bf R}) \cdot d{\bf r}_2 = \ell\,, 
\end{equation}
where the contour integral is taken along a circuit enclosing the vortex center. These relations show that the center of the first-order $|\ell |=1$ WWV can be considered as the first-order phase singularity in the scalar field $Z(x,y)$ or the second-order polarization singularity (C-point of circular polarization) in the vector field ${\bf R}(x,y)$ \cite{Nye1987,BerryDennis2001,BAD2019,Bliokh2021POF}. Any perturbation breaking the cylindrical symmetry splits the second-order C-point into a pair of the first-order C-points, with topologically-robust M\"{o}bius-strip orientations of the polarization ellipses around these points \cite{Freund2010,Bauer2015,BAD2019,Bliokh2021POF,Muelas2022}. 

Nonzero vorticity and circulation do appear in WWVs, but in the {\it quadratic} corrections to linear wave solutions. Namely, water particles experience a slow {\it Stokes drift}, i.e., the difference between the Lagrangian and Euler velocities \cite{Bremer2017,Falkovich_book,Bliokh2022}:
\begin{equation}
\label{Stokes}
{\bf U} = \frac{\omega}{2} {\rm Im} [{\bf R}^* \cdot ({\bm \nabla}_2) {\bf R}]\,.
\end{equation}
Multiplied by the mass density, it yields the canonical {\it wave momentum} (`pseudomomentum')  density \cite{Bliokh2022,McIntyre1981,Peierls_I,Peierls_II,Bliokh2022PRA}: ${\bf P} = \rho {\bf U}$.  

Figure~\ref{Fig_1}(c) shows the azimuthal Stokes-drift flow in WWVs. It is mostly localized near the first radial maximum of the Bessel function $J_\ell (kr)$ and determines the $z$-directed OAM density: ${\bf L} = {\bf r}_2 \times {\bf P}$, $L_z = (\rho\omega/2) {\rm Im} ({\bf R}^* \cdot \partial_\varphi {\bf R})$. Notably, the local circular motion of water particles (spin) and the global Stokes-drift circulation (OAM) have very different space and time scales: the linear-wave amplitude $A$ and angular frequency $\omega$ vs. the wavelength $k^{-1} \gg A$ and angular velocity $U/r \sim \omega k^2 A^2 \ll \omega$. The spin and OAM densities in the WWVs (\ref{Z_Bessel}) and (\ref{a_Bessel}) satisfy the relation following from Eq.~(\ref{Jz}) \cite{Bliokh2019_II,Bliokh2022PRL}:
\begin{equation}
\label{Jz_density}
J_z = L_z + S_z = \frac{\rho \omega}{2} \ell |{\bf R}|^2 = 2 \frac{\ell}{\omega} T\,,
\end{equation}
where $T = \rho |{\bf V}|^2/4$ is the cycle-averaged kinetic energy density. 

Thus, WWVs are naturally described by a quantum-like formalism and possess nontrivial topological properties. 
Recent experiments \cite{Filatov2016,Filatov2016_II,Francois2017,Bliokh2022} generated square lattices of alternating first-order vortices with $\ell = \pm 1$ by interfering orthogonal standing waves with the $\pi/2$ phase difference. The orbital Stokes drift and circular polarization (spin) in the vortex centers were clearly observed in Refs.~\cite{Francois2017,Bliokh2022}, but quantized topological properties of these vortices have not been described. Higher-order WWVs with $|\ell | >1$, which have never been observed,  could provide areas of unperturbed water surface surrounded by intense circular waves and orbital Stokes flows.

\begin{figure*}[th]
\centering
\includegraphics[width=0.9\linewidth]{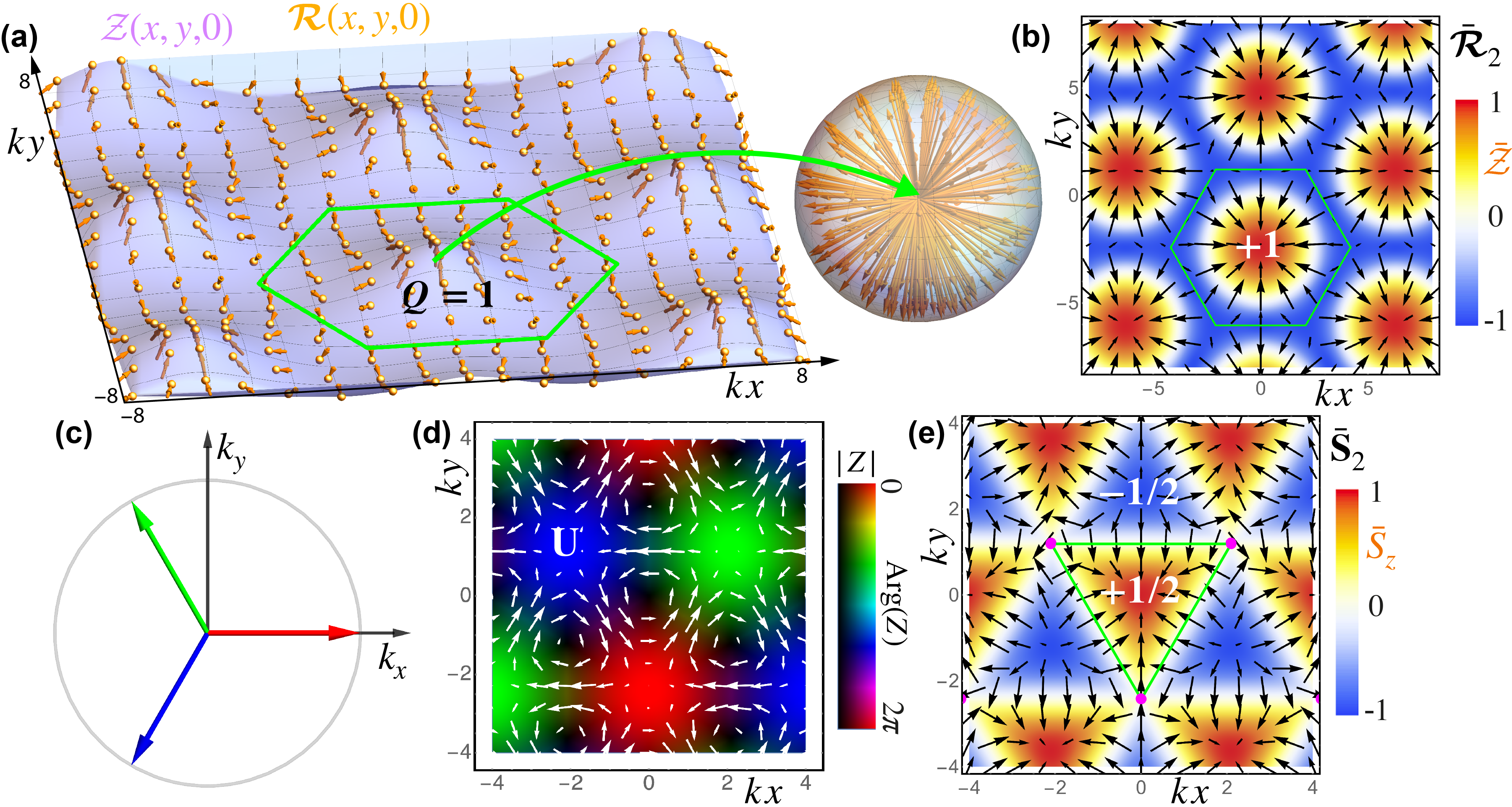}
\caption{Hexagonal lattice produced by the interference of three waves with equal frequencies, amplitudes, and color-coded phases shown in (c). (a) Instantaneous water surface ${\mathcal Z}(x,y,0)$ and water-surface particle displacements $\mathbfcal{R}(x,y,0)$ for the field (\ref{eq:system}). The displacement directions in the unit hexagonal cell is mapped onto the unit sphere, providing a skyrmion with the topological charge $Q=1$, Eq.~(\ref{skyrmion}). (b) The unit displacement-direction field $\bar{\mathbfcal{R}}(x,y,0)$ represented by colors (vertical component $\bar{\mathcal{Z}}$) and black arrows (in-plane components $\bar{\mathbfcal{R}}_2$). 
(d) The complex vertical-displacement field $Z(x,y)$ and the Stokes drift ${\bf U}$ indicating the lattice of alternating WWVs with $\ell=\pm 1$. 
(e) The unit spin-density field $\bar{\bf{S}}(x,y)$ represented similar to (b). The hexagonal unit cell is split into triangular zones of spin merons (half-skyrmions) with topological charges $Q_S = \pm 1/2$ and centers with $\bar{S}_z = \pm 1$ corresponding to the $\ell=\pm 1$ vortices in (d). }
\label{Fig_2}
\end{figure*}

{\it Water-wave skyrmions and merons.---}The 3D vector nature of water waves allows the generation of topological vector textures, such as skyrmions or merons \cite{Nagaosa2013,Tsesses2018,Du2019,Dai2020,Shen2021,Lei2021,Deng2022,Ge2021,Muelas2022,Cao2023}. Such textures can be produced by interfering several plane waves with the same frequency and wavevectors ${\bf k}_j = k(\cos{\phi}_j, \sin{\phi_j},0)$, $j=1,...,N$:
\begin{equation} 
\label{Interference}
{\bf R}= \sum_{j=1}^N {\bf R}_{0j} e^{{i{\bf k}_j\cdot {\bf r} + i\Phi_j}},~
{\bf R}_{0j} = A_{j} (i\cos{\phi}_j, i \sin{\phi_j}, 1),
\end{equation}
where $A_{j}$ and $\Phi_{j}$ are the real-valued amplitudes and phases of the interfering waves. 

Consider, for example, $N=3$ waves, uniformly distributed with $\phi_j=2\pi(j-1)/N$, $A_{j}=A$, and vortex phases $\Phi_j=\phi_j$, Fig.~\ref{Fig_2}(c). These waves form a hexagonal periodic lattice with the displacement field 
\begin{equation} 
\label{eq:system}
\begin{pmatrix}
X \\  Y \\  Z \\ 
\end{pmatrix} \propto A\! 
\begin{pmatrix}
i e^{ik{x}} + i e^{-i\frac{k{x}}{2}} \sin\!\left(\frac{\sqrt{3} k{y}}{2}+\frac{\pi}{6}\right) \\
- \sqrt{3} e^{-i\frac{k{x}}{2}} \cos\!\left(\frac{\sqrt{3} k{y}}{2}+\frac{\pi}{6}\right) \\
e^{ik{x}} - 2 e^{-i\frac{k{x}}{2}} \sin\!\left(\frac{\sqrt{3} k{y}}{2}+\frac{\pi}{6}\right)
\end{pmatrix}\!.
\end{equation}
This field exhibits a number of nontrivial topological features.
First, it contains a lattice of WWVs with alternating topological charges $\ell = \pm 1$, Fig.~\ref{Fig_2}(d). Such vortex lattices are well known in optics \cite{Masajada2001}.

Second, Fig.~\ref{Fig_2}(a) shows the instantaneous water surface ${\mathcal Z}({\bf r}_2,0)$ and the surface-particle displacements $\mathbfcal{R}({\bf r}_2,0)$ for the field (\ref{eq:system}). The displacements in a hexagonal unit cell of the lattice contain all possible directions and can be mapped onto a unit sphere. This is a signature of a skyrmion, which can be characterized by the topological number 
\begin{equation} 
\label{skyrmion}
Q =\frac{1}{4 \pi} \iint_{\text{u.c.}} \bar{\mathbfcal{R}} \cdot [\partial_x \bar{\mathbfcal{R}} \times \partial_y \bar{\mathbfcal{R}}] \, dx \, dy,
\end{equation}
where $ \bar{\mathbfcal{R}} =  {\mathbfcal{R}}/ | {\mathbfcal{R}}|$. In the case under consideration, $Q=1$ at $t=0$, but it can change its sign over time, because the displacement evolves and becomes opposite after half a period, $t=\pi/\omega$ \cite{Muelas2022}. Figure~\ref{Fig_2}(b) displays another representation of the skyrmion lattice, where colors and black vectors indicate the $z$ and $(x,y)$ components of the displacement-direction field $\bar{\mathbfcal{R}}$. 
Moving from the center of the cell towards its boundary, the vector $\bar{\mathbfcal{R}}$ undergoes a rotation, where its $z$-component changes sign, resulting in a nontrivial winding captured by the nonzero skyrmion charge $Q$. Similar skyrmion lattices have been observed in electromagnetic \cite{Tsesses2018}, sound \cite{Ge2021,Muelas2022}, and elastic \cite{Cao2023} vector wavefields.

Third, instead of the instantaneous vector field ${\mathbfcal{R}}$, one can trace the spin density vector ${\bf S}$ (normal to the local polarization ellipse). Figure~\ref{Fig_2}(e) displays the distribution of the unit spin vector $\bar{\bf S} = {\bf S}/|{\bf S}|$ in the field (\ref{eq:system}). The unit hexagonal cell is split into triangular zones with $\bar{S}_z > 0$ and $\bar{S}_z < 0$ separated by $\bar{S}_z = 0$ lines and singular ${\bf S} = {\bf 0}$ vertices. The centers of these triangles with $\bar{S}_z = \pm 1$ (i.e., circular in-plane polarizations) correspond to the centers of WWVs with $\ell=\pm 1$, Fig.~\ref{Fig_2}(d) \cite{Bliokh2022}. Calculating the topological charges (\ref{skyrmion}) for the spin field $\bar{\bf S}$, we obtain $Q_S=\mp 1/2$ for the triangular zones with $\bar{S}_z \lessgtr 0$. Such topologically nontrivial textures are called {\it merons} or half-skyrmions, because the spin directions in each zone covers the upper or lower semisphere. Similar spin merons have been observed in electromagnetic waves \cite{Dai2020,Lei2021,Ghosh2021,Krol2021}.

Here we showed only one simple example of the water-wave interference field. WWVs, field skyrmions, and spin merons are rather universal topological entities and appear in many other fields. A square lattice formed by two standing waves \cite{Filatov2016,Filatov2016_II,Francois2017,Bliokh2022} contains vortices and spin merons (cf. \cite{Lei2021,Ghosh2021}), \blue{a hexagonal lattice formed by three standing waves produces field skyrmions \cite{SM}}, and the zero-order $\ell=0$ Bessel mode, Eqs.~(\ref{Z_Bessel}) and (\ref{a_Bessel}) and Fig.~\ref{Fig_1}, contains a field skyrmion (cf. \cite{Tsesses2018,Deng2022}). 

\begin{figure}[t!]
\centering
\includegraphics[width=\linewidth]{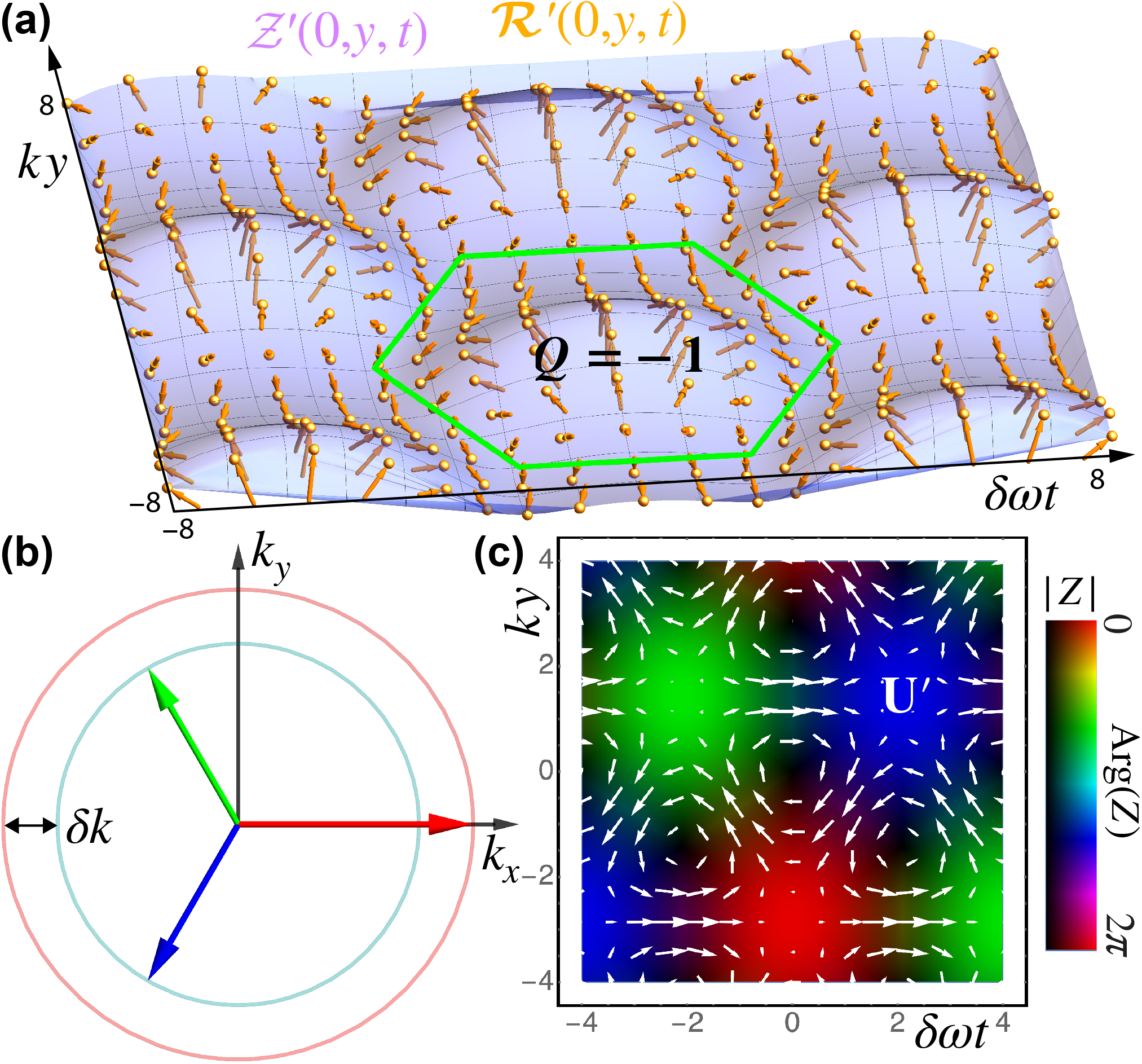}
\caption{Same as in Figs.~\ref{Fig_2}(a,c,d) but for the lattice of spatiotemporal WWVs and skyrmions. The frequency of one of the interfering waves is detuned by $\delta \omega$ and the corresponding $\delta k$. The complex vertical-displacement field $Z$ and the real field envelope ${\mathbfcal R}'  = {\rm Re}[{\bf R}]$ (without fast oscillations $e^{-i\omega t}$) are plotted over the spacetime domain $(t,y)$ at the fixed coordinate $x=0$. The temporal component of the `spatiotemporal Stokes drift' ${\bf U}' = (U_t, U_y)$ is defined as 
$U_t = (k{\omega}/{2}\delta\omega) {\rm Im} [{\bf R}^* \cdot (\partial_t) {\bf R}]$.}
\label{Fig_3}
\end{figure}

{\it Spatiotemporal vortices and skyrmions.---}Finally, we demonstrate another class of topological entities which can be readily generated in water waves: {\it spatiotemporal} vortices \cite{Sukhorukov2005,Bliokh2012,Hancock2019,Chong2020,Bliokh2021} and skyrmions. 
It is sufficient to slightly detune the frequency of one of the three interfering plane waves in Fig.~\ref{Fig_2}: $\omega_1 \to \omega + \delta\omega$, $k_1 \to k + \delta k = (\omega+\delta\omega)^2/g$ \blue{(for simplicity, here we neglect capillarity, $\alpha\to 0$)}, Fig.~\ref{Fig_3}(b). This transforms the wavefield (\ref{Interference}) as $\Phi_1 \to \Phi_1 - i\delta\omega t$, so that the spatial lattice in Fig.~\ref{Fig_2} becomes {\it moving} along the $x$-axis, and the field becomes a function of space and {\it time}: ${\bf R}({\bf r}_2,t)$.

The real displacement field is $\mathbfcal{R}({\bf r}_2,t) = {\rm Re}[{\bf R}({\bf r}_2,t)e^{-i\omega t}]$, but we will analyze the field $\mathbfcal{R}' ({\bf r}_2,t) = {\rm Re}[{\bf R}({\bf r}_2,t)]$ subtracting the common fast oscillations $e^{-i\omega t}$.
Plotting the complex field $Z$ and real field $\mathbfcal{R}'$ in the spacetime domain $(t,y)$ at fixed  $x=0$, we find that they exhibit a scaled hexagonal lattice of vortices and skyrmions, Fig.~\ref{Fig_3}. These spatiotemporal WWVs and skyrmions have opposite topological charges $\ell$  and $Q$ compared to their spatial counterparts in Fig.~\ref{Fig_2}.

{\it Conclusions.---}We have analyzed the fundamental topologically nontrivial objects in linear water-surface (\blue{gravity-capillary}) waves, namely: WWVs, surface-particle displacement skyrmions, spin-density merons, as well as spatiotemporal WWVs and skyrmions. All these objects are universal across different types of waves and only require standard wave-interference ingredients: relative phases/amplitudes, polarizations, and spectral detuning, to control the geometry and topology of the field. 
\blue{For simplicity, we considered the deep-water approximation, the finite-depth effects in monochromatic water waves simply produce global scaling of the vertical component on the surface: $Z \to \tanh(kH) Z$, where $H$ is the water depth \cite{SM}.}

Notably, the vector features of water waves (displacement fields) are directly observable, while in other fields these are usually measured via various indirect methods. Therefore, water waves offer a highly attractive platform for emulating topologically nontrivial field structures and wave phenomena in a unified fashion. Furthermore, nontrivial dynamical properties of topological water-wave objects --- circulating Stokes-drift currents, fast circular motions (spin) in the centers of the first-order WWVs, vanishing fields in the centers of higher-order WWVs, etc. --- can be attractive for fluid-mechanical applications, such as manipulations of particles \cite{Ding2013,Wu2019}.  
\blue{Finally, we note that while most of the attention in water-wave physics has focused on nonlinear and high-amplitude effects \cite{Cazaubiel2019, Birkholz2016, McAllister2022}, our study shows that wave structures around field zeros and linear-wave interference exhibit a rich variety of largely unexplored phenomena.}   

\begin{acknowledgments}
This work is supported in part by
the Japan Society for the Promotion of Science (JSPS);
Nippon Telegraph and Telephone Corporation (NTT) Research;
the Asian Office of Aerospace Research and Development (AOARD) [Grant No. FA2386-20-1-4069];
the Foundational Questions Institute Fund (FQXi) [Grant No. FQXi-IAF19-06];  the International Research Agendas Programme (IRAP) of the Foundation for Polish Science co-financed by the European Union under the European Regional Development Fund and Teaming Horizon 2020 programme of the European Commission [ENSEMBLE3 Project MAB/2020/14]; the TEAM program of the Foundation for Polish Science co-financed by the European Union under the European Regional Development Fund [Grant TEAM/2016-3/29]; and the Australian Research Council [FT230100058].
\end{acknowledgments}

\bibliography{Vortex-skyrmion_1}

\end{document}